\documentstyle[preprint,aps]{revtex}
\tightenlines
\newcommand {\be}{\begin{equation}}
\newcommand {\ee}{\end{equation}}
\newcommand{\bey}{\begin{eqnarray}}
\newcommand{\eey}{\end{eqnarray}}

\begin{document}
\title{Classical Representation of the 1D Anderson Model}
\author {F.M. Izrailev$^{1,2}$, S. Ruffo$^{3}$, and L. Tessieri$^{4}$} 
\address{
$^1$ Instituto de Fisica, Universidad Autonoma de Puebla, \\
Apdo. Postal J-48, Col. San Manuel, Puebla, Pue. 72570, Mexico\\
$^2$ Budker Institute of Nuclear Physics, Novosibirsk 630090, Russia\\
$^3$ Dipartimento di Energetica "Sergio Stecco", Universit\`a degli Studi di 
Firenze, Via di Santa Marta, 3 50139 Firenze, Italy, and INFN, Firenze.\\
$^4$ Dipartimento di Fisica, Universit\`a degli Studi di Firenze, Largo E.
Fermi, 2 50125, Firenze, Italy, and INFN, Firenze.
}
\date{\today}
\maketitle

\begin{abstract}
A new approach is applied to the 1D Anderson model by making use of a
two-dimensional Hamiltonian map. For a weak disorder this approach allows
for a simple derivation of correct expressions for the localization length
both at the center and at the edge of the energy band, where standard
perturbation theory fails. Approximate analytical expressions for strong
disorder are also obtained.
\end{abstract}



\newpage

\section{Introduction}

Recently, it was suggested to treat 1D tight-binding models with diagonal
disorder in terms of classical Hamiltonian maps~\cite{Izr95a}. This approach
has been successfully used in the description of delocalized states in the
so-called  dimer model~\cite{Izr95b}, as well as for the Kronig-Penney model~%
\cite{Izr97}. In this paper we show that even for the standard 1D Anderson
model, this approach allows us to obtain new analytical results and
reproduce known results in a much more transparent way, by making reference
to the properties of the dynamics of the noisy Hamiltonian map into which
the model is transformed.

As was indicated in~\cite{Izr95a,Izr95b}, the discrete stationary
Schr\"odinger equation 
\begin{equation}
\label{Schro}\psi _{n+1}+\psi _{n-1}=(\epsilon _n+E)\psi _n~,
\end{equation}
with $\epsilon _n$ standing for the diagonal potential and $E$ for the energy
of an eigenstate, can be written in the form of a two-dimensional
Hamiltonian map 
\begin{eqnarray}x_{n+1} &=& x_n \cos \mu - (p_n + A_n x_n) \sin \mu \nonumber \\
p_{n+1} &=& x_n \sin \mu + (p_n + A_n x_n) \cos \mu~.
\label{Ham}
\end{eqnarray}Here, the variables $(p_n,x_n)$ play the role of the momentum
and position of a linear oscillator subjected to linear periodic delta-kicks
with the period $T=1$. The amplitude $A_n$ of the kicks depends on time
according to the relation $A_n=-\epsilon _n/\sin \mu $. For the Anderson
model the distribution $P(\epsilon )$ of the disorder is given by $%
P(\epsilon )=1/W$ for $|\epsilon |\leq W/2$, with variance $\langle
\epsilon ^2\rangle =\sigma ^2=W^2/12$. Between two successive kicks, the
rotation in the phase space is given by the eigenstate energy, $E=2\cos \mu $%
. In such a representation, amplitudes $\psi _n$ of a specific eigenstate at
site $n$ correspond to positions of the oscillator at times $t_n=n$ and,
therefore, the structure of eigenstates can be studied by investigating the
time-dependence of the trajectories in the phase space $(p_n,x_n)$. In
particular, localized states correspond to unbounded trajectories and,
vice-versa, extended states are represented by bounded trajectories.

It is convenient to pass to action-angle variables $(r_n,\theta _n)$
according to the standard transformation, $x=r\sin \theta ,\,p=r\cos \theta $%
. The corresponding map, therefore, has the form 
\begin{eqnarray}
\label{Polar}
r_{n+1} &=& r_n D_n \nonumber \\
\sin \theta_{n+1} &=& D_n^{-1} \left(\sin (\theta_n - \mu) - A_n \sin \theta_n
\sin \mu \right) \\
\cos \theta_{n+1} &=& D_n^{-1} \left(\cos (\theta_n - \mu) + 
A_n \sin \theta_n \cos \mu \right)~, \nonumber
\end{eqnarray}
where 
\begin{equation}
\label{Dienne}D_n=\sqrt{1+A_n\sin (2\theta _n)+A_n^2\sin {}^2\theta _n}~.
\end{equation}
The localization length $l$ is defined by the standard relation 
\begin{equation}
\label{Loc}l^{-1}=\lim _{N\to \infty }\langle \frac 1N\sum_{n=0}^{N-1}\ln
\left| \frac{x_{n+1}}{x_n}\right| \rangle =\overline{\langle \ln \left| 
\frac{x_{n+1}}{x_n}\right| \rangle }~,
\end{equation}
where the overbar stays for time average and the brackets for the average
over different disorder realizations. The contributions to $l^{-1}$ can be
splitted in two terms 
\begin{equation}
\label{loc}l^{-1}=\overline{\langle \ln \left( \frac{r_{n+1}}{r_n}\right)
\rangle }+\overline{\langle \ln \left| \frac{\sin \,\theta _{n+1}}{\sin
\,\theta _n}\right| \rangle }~.
\end{equation}
The second term on the r.h.s. is negligible because it is the average of a
bounded quantity. It becomes important only when also the first term is
small, i.e. at the band edge $\mu \approx 0$. Thus, apart from this limit,
the localization length can be evaluated from the map (\ref{Polar}) using
only the dependence of the radius $r_n$ on discrete time. The ratio $%
r_{n+1}/r_n$ is a function only of the angle $\theta _n$ and not of the
radius $r_n$, thus the computation of the localization length implies just
the average over the invariant measure $\rho (\theta )$, which is an
advantage with respect to transfer matrix methods. Moreover, since $%
r_{n+1}/r_n$ is positive, there is no need to work with complex quantities.

In a direct analytical evaluation of (\ref{Loc}) one can, therefore, write 
\begin{equation}
\label{Measure}l^{-1}=\int P(\epsilon )\int_0^{2\pi }\ln (D(\epsilon ,\theta
))\,\rho (\theta )\,{d\theta }{d\epsilon }~,
\end{equation}
where $P(\epsilon )$ is the density of the (uncorrelated) distribution of $%
\epsilon _n$, and $\rho (\theta )$ stands for the invariant measure of the
one-dimensional map for the phase $\theta $, see (\ref{Polar}). We use here
the fact that $\rho (\theta )$ does not depend on the specific sequence $%
\epsilon _n$, but can depend on the moments of $P(\epsilon )$, particularly
on its second moment $\sigma ^2$ (see below). As one can see, the main
problem is in the expression for $\rho (\theta )$, which was not found
explicitly even in the limit of a weak disorder, $A_n\to 0$~\cite
{Kap81,Der84}.

\section{Weak disorder}

By weak disorder we mean that $A_n$ is small. This can be arranged even at
the band-edge, where the denominator $\sin \,\mu $ of $A_n$ is also small;
thus the disorder $\epsilon _n$ must go to zero faster than $\mu $ (how much
faster, it is determined by the properties of the Hamiltonian map).
Retaining only terms up to $O(A_n^2)$ in the map (\ref{Polar}) for $\theta _n
$ one gets 
\begin{equation}
\label{smallmap}\theta _{n+1}=\theta _n-\mu -A_n\sin {}^2\theta _n+A_n^2\sin
{}^3\theta _n\cos \,\theta _n\,\,\,\,\,\,\,\,\,\,\,\,\,\,\,\mbox{mod}%
\,\,2\pi ~,
\end{equation}
which coincides with formula (62) in Ref.~\cite{Kap81}.

The expression (\ref{Measure}) for $l^{-1}$ can be written in the weak
disorder limit explicitly, 
\begin{equation}
\label{smalllyap}l^{-1}=\frac{1}{2\sin {}^2\mu }\int \epsilon ^2 P(\epsilon
)\,d\epsilon \int\limits_0^{2\pi }\rho (\theta )\left( \frac 14-\frac 12\cos
(2\theta )+\frac 14\cos (4\theta )\right) d{\theta }~,
\end{equation}
which is valid over all the spectrum except at the band edge, where the
additional contribution in (\ref{loc}) is present (see below). In fact,
standard perturbation theory~\cite{Tho79} corresponds to the assumption that 
$\rho (\theta )$ is constant. Thus, one easily obtains 
\begin{equation}
\label{standard}l^{-1}=\frac{\sigma ^2}{8\sin {}^2\mu }=\frac{W^2}{96\left(
1-\frac{E^2}4\right) }~.
\end{equation}
This expression was found to work quite well over all energies, but,
surprisingly, numerical experiments~\cite{Czy81} showed a $4\%$ deviation at
the band center. One had to explain why standard perturbation theory
fails at the band center, while it is correct everywhere else. Non-standard
perturbation theory methods were devised in Refs.~\cite{Kap81,Der84}, where
the correct value for $l^{-1}$ at the band center was obtained and,
moreover, a different scaling with disorder was discovered at the band-edge~%
\cite{Der84}. However, these methods hide behind mathematical difficulties
the physical origin of the discrepancy. We are here able, by looking at the
properties of map (\ref{smallmap}), to understand both the physical nature
of the discrepancy at the band center and the different scaling at the band
edge. Moreover, our derivation is much simpler mathematically and more
straightforward (this can be already seen from the very simple derivation of
the expression (\ref{standard})).

\subsection{The band center}

In order to derive analytically the correct expression for $l^{-1}$ at the
band center $E=0$ , one has to find the exact expression for the invariant
probability measure $\rho (\theta )$. This latter arises from map (\ref
{smallmap}) specialized to the value $\mu =\pi /2$. For vanishing disorder
the trajectory is a period four, specified by the initial angle $\theta _0$.
For a weak disorder any orbit diffuses around the period four, with an
additional drift in $\theta $. Asymptotically, any initial condition gives
rise to the same invariant distribution, which can now be expected to be
different from constant. To find this distribution, we write the fourth
iterate of the map (\ref{smallmap}) 
\begin{equation}
\label{iterate}\theta _{n+4}=\theta _n-\xi _n^{(1)}\sin {}^2\theta _n-\xi
_n^{(2)}\cos {}^2\theta _n-\frac{\sigma ^2}2\sin \,(4\theta _n)~,
\end{equation}
where $\xi _n^{(1)}=\epsilon _n+\epsilon _{n+2}$ and $\xi _n^{(2)}=\epsilon
_{n+1}+\epsilon _{n+3}$ are uncorrelated random variables with zero mean and
variance $2\sigma ^2$. Here, we have neglected in Eq.~(\ref{iterate}) mixed
terms of the kind $\epsilon _n\epsilon _m$ ($m\neq n$) and approximated $%
\langle (\xi _n^{(1)})^2\rangle $ and $\langle (\xi _n^{(2)})^2\rangle $ by
their common variance $2\sigma ^2$, which is meaningful in a perturbative
calculation at first order in $\epsilon _n$.

Thus, the invariant distribution can be determined analytically in the
continuum limit where $\theta _{n+4}-\theta _n$ is replaced with $d\theta $
and the random variables $\xi _n^{(1)},\xi _n^{(2)}$ with the Wiener
variables $dW_1,dW_2$ with properties 
\begin{eqnarray*}
\langle dW_i \rangle &=& 0 \\
\langle dW_i dW_j \rangle &=& 2 \delta_{ij} \sigma^2 dt \;\;i,j=1,2
\end{eqnarray*}
obtaining the Ito equation 
\begin{equation}
d\theta =-dW_1\sin {}^2\theta -dW_2\cos {}^2\theta -\frac{\sigma ^2}2\sin
\,(4\theta )\,dt~.
\end{equation}
To this we can associate the Fokker-Planck equation~\cite{Gardiner} 
\begin{equation}
\label{FP}\frac{\partial P}{\partial t}(\theta ,t)=\frac{\sigma ^2}2\frac
\partial {\partial \theta }\left( \sin (4\theta )P(\theta ,t)\right) +\frac{%
\sigma ^2}4\frac{\partial ^2}{\partial \theta ^2}\left[ (3+\cos (4\theta
))P(\theta ,t)\right] ~.
\end{equation}
The stationary solution $\rho (\theta )$ of Eq.~(\ref{FP}), satisfying the
conditions of periodicity $\rho (0)=\rho (2\pi )$ and normalization $%
\int_0^{2\pi }\rho (\theta )=1$, is 
\begin{equation}
\label{stationary}\rho (\theta )=\left( 2{\bf K}\left( \frac 1{\sqrt{2}%
}\right) \,\sqrt{3+\cos (4\theta )}\right) ^{-1}~,
\end{equation}
where ${\bf K}$ is the complete elliptic integral of the first kind. One
should note that the expression for the invariant measure has never been
derived before, and it could turn out to be useful for obtaining observables
other than $l^{-1}$. Note that solution (\ref{stationary}) does not depend
on the strength of the random process.

Inserting formula (\ref{stationary}) into (\ref{smalllyap}) at $\mu =\pi /2$
we get 
\begin{equation}
\label{elle}l^{-1}=\frac{\sigma ^2}8\left( 1+\int\limits_0^{2\pi }\rho
(\theta )\cos (4\theta )\,d\theta \right) =\sigma ^2\left( \frac{\Gamma (3/4)
}{\Gamma (1/4)}\right) ^2=\frac{W^2}{105.2\dots }
\end{equation}

This result perfectly agrees with the one obtained in Ref.~\cite{Der84,Econo}%
, although it is here derived with a different approach, that involves much
simpler calculations.

Thouless' standard perturbation theory result would correspond to neglect
the average of the $\cos (4\theta )$ term in (\ref{elle}), meaning that the
stationary solution (\ref{stationary}) is approximated with a flat
distribution. This approximation works well for all energies $E=2\cos \,\mu $%
, with $\mu =\alpha \pi $ and $\alpha $ irrational, but doesn't work for the
band-center. Moreover, if one would consider observables which contain
higher harmonics than those present in the formula for $l^{-1}$ (\ref
{smalllyap}), one would get corrections to standard perturbation theory also
for other rational values of $\alpha =p/q$. In fact, numerical experiments
show that the invariant measure for rationals is modulated with the main
period $T=\pi/q$ ($p$ and $q$ being prime to each other). The amplitude 
of the modulation decreases with $q$; thus the strongest modification is obtained
for $\alpha =1/2$, which corresponds to the band-center. It is moreover
clear from formula (\ref{smallmap}), that the only energy value for
which a contribution due to the modulations in the measure $\rho (\theta )$
is present in the inverse localization length $l^{-1}$ is the band center $%
\alpha =1/2$. This is because in Eq.~(\ref{smalllyap}) only second and
fourth order harmonics must be averaged, and only for $\alpha =1/2$ the
fourth harmonic occurs in $\rho (\theta )$. This is of course true only in
the small disorder limit.

\subsection{The band edge}

The neighbourhood of the band edge corresponds to $\mu \approx 0$. If the
second order noisy term $A_n^2$ in the map (\ref{smallmap}) is replaced by
its average, which is the same approximation we did in the previous Section,
the map (\ref{smallmap}) reduces to 
\begin{equation}
\label{bandmap}\theta _{n+1}=\theta _n-\mu +\frac{\epsilon _n}\mu \sin
{}^2\theta _n+\frac{\delta ^2}{\mu ^2}\sin {}^3\theta _n\cos \,\theta
_n\,\,\,\,\,\,\,\,\,\mbox{mod}\,\,2\pi ~,
\end{equation}
where $\delta ^2$ is the variance of the noise $\epsilon _n$. For vanishing
disorder and $\mu \to 0$ the orbits are fixed points. Moving away from the
band edge produces a quasi-periodic motion and switching on the disorder
gives rise to diffusion. Following the procedure of the previous Section
(but here we do not have to go to the four-step map), we obtain the
corresponding Fokker-Planck equation 
\begin{equation}
\label{FP2}\frac{\partial P}{\partial t}(\theta ,t)=\frac \partial {\partial
\theta }\left( \mu -\frac{\delta ^2}{\mu ^2}\sin {}^3 \theta \cos \theta
P(\theta ,t)\right) +\frac{\delta ^2}{2\mu ^2}\frac{\partial ^2}{\partial
\theta ^2}\left( \sin {}^4\theta \,P(\theta ,t)\right) ~.
\end{equation}
There are in this case two small quantities: the noise $\epsilon _n$ and the
distance from the band edge $\Delta =2-2\cos \,\mu \approx \mu ^2$. Below we
consider the double limit $\Delta \to 0$, $\delta ^2\to 0$. One can see
that, if we keep the ratio $k=\mu ^3/\delta ^2$ fixed, the time scale of the
drift term in (\ref{FP2}) is unique and, moreover, it coincides with the
diffusion time scale, being $1/\mu $. We can thus rescale time $\tau =t\mu $
and obtain the following stationary Fokker-Planck equation, 
\begin{equation}
\label{FP3}\frac \partial {\partial \theta }\left( k-\sin {}^3\theta \cos
\theta \,\rho (\theta )\right) +\frac 12\frac{\partial ^2}{\partial \theta
^2}\left( \sin {}^4\theta \,\rho (\theta )\right) =0~,
\end{equation}
which depends only on $k\,$. Its solution, with the same normalization and
periodicity conditions as above, is 
\begin{equation}
\rho (\theta )=\frac{f(\theta )}{\sin {}^2\theta }\left[
C+\int\limits_0^\theta dx\frac{2J}{f(x)\sin {}^2 x }\right] ~,
\end{equation}
where 
\begin{equation}
f(\theta )=\exp \left( 2k\left( \frac 13\cot {}^3\theta +\cot \,\theta
\right) \right) ~,
\end{equation}
and $C,J$ are integration constants. To make $\rho $ normalizable, constant $%
C$ must vanish and $J$ is then fixed by the normalization condition, 
\begin{equation}
J^{-1}=\frac{\sqrt{8\pi }}{k^{2/3}}\int\limits_0^\infty dx\frac 1{\sqrt{x}%
}\exp \left( -\frac{x^3}6-2k^{2/3}x\right) ~.
\end{equation}
As was mentioned in Section I, in the evaluation of the inverse localization
length given by (\ref{loc}) we must now take into account both terms on the
r.h.s.; thus, we come to the expression 
\begin{equation}
l^{-1}=\overline{\left\langle \ln \left| D_n\frac{\sin \theta _{n+1}}{\sin
\theta _n}\right| \right\rangle }.
\end{equation}
In the limit of a weak disorder and for $\mu \to 0$ one gets 
\begin{equation}
l^{-1}=-\mu \,\overline{\langle \cot \,\theta _n\rangle }=-2\mu
\int\limits_0^\pi \cot \,\theta \,\rho (\theta )\,d\theta ~.
\end{equation}
After some straightforward calculations, with $\mu =(k\delta ^2)^{1/3}$, one
obtains 
\begin{equation}
l^{-1}=\frac{(\delta ^2)^{1/3}}2\frac{\int_0^\infty dx\,x^{1/2}\exp \left( -
\frac{x^3}6-2k^{2/3}x\right) }{\int_0^\infty dx\,x^{-1/2}\exp \left( -\frac{%
x^3}6-2k^{2/3}x\right) }~,
\end{equation}
which coincides with expression (36) in Ref.~\cite{Der84}. The limits $%
k\to 0$ and $k\to \infty $ are then easily rederived and coincide with those
in Ref.~\cite{Der84}. For instance, the $k\to 0$ limit gives the scaling law 
\begin{equation}
l^{-1}=\frac{6^{1/3}\sqrt{\pi }}{2\Gamma \left( \frac 16\right) }(\delta
^2)^{1/3}=0.289\dots (\delta ^2)^{1/3}~.
\end{equation}
It is interesting to observe that a similar scaling law was also found for
chaotic billiards (stadia and oval ones) looking at the behavior of the
Lyapunov exponent in the integrable limit~\cite{Ben84}. It is quite natural
to associate the Lyapunov exponent with the inverse localization length, and
the geometrical parameter that in billiard measure the distance from
integrability, with the intensity of the disorder $\sqrt{\delta ^2}$ in the
Anderson model.

We have seen in this Section that the study of the dynamics of the noisy
circle map (\ref{smallmap}) allows us to derive both the standard Thouless
perturbation theory results for the behavior of the localization length in
the weak disorder limit, and the non-standard corrections to such a theory
both at the band-center and at the band edge. The advantage of our method is
that the derivation of the final formula has a clear physical meaning;
moreover, for the band-center case, the procedure is mathematically more
straightforward than those previously used~\cite{Kap81,Der84}.

\section{ Strong disorder}

As is known, the analytical expression for the localization length was found
only in the limiting cases of a very weak or a very strong disorder. It is
interesting that relation (\ref{Measure}) allows us to derive an
approximate expression which is also good for a quite large disorder.
Indeed, if the energy is not close to the band edge and the disorder is not
very large, one can expect a strong rotation of the phase $\theta $ .
Therefore, the invariant measure $\rho (\theta )$ can be approximately taken
as constant, $\rho (\theta )=\left( 2\pi \right) ^{-1}$ . For such a
disorder, one can explicitly integrate Eq.(\ref{Measure}), first over the
phase $\theta $, 
\begin{equation}
\label{theta}\frac 1{4\pi }\int\limits_0^{2\pi }\ln \left( 1+A\sin (2\theta
)+A^2\cos {}^2\theta \right) d\theta =\frac 12\ln \left( 1+\frac{A^2}%
4\right) ;\,\,\,\,\,\,\,\,A^2=\epsilon ^2/\sin {}^2\mu \, 
\end{equation}
and after, over the disorder $\epsilon $, 
\begin{equation}
\label{l1}l_w^{-1}=\frac 12 \int P(\epsilon )\ln 
\left( 1+\frac{\epsilon ^2}{4\sin
{}^2\mu }\right) d\epsilon =\frac 12\ln \left( 1+\frac{W^2}{4\sin {}^2\mu }%
\right) +\frac{Arctg\left( \frac W{2\sin \mu }\right) }{\frac W{2\sin \mu }}%
-1 
\end{equation}
Direct numerical simulations show that this expression gives quite a
good agreement with the data for the disorder $W\leq 1\div 3$ inside the
energy range $\left| E\right| \leq 1.85$ .

For a much stronger disorder, one can use another approach. Note, that for
the unstable region 
\begin{equation}
\label{unstable}|E-\epsilon _n|>\,2 
\end{equation}
of the one-step Hamiltonian map (\ref{Ham}) both eigenvalues 
$\lambda _n^{(1,2)}$ are real,
\begin{equation}
\label{eigen}\lambda _n^{(1,2)}=\frac 12\left( (E-\epsilon _n)\,\pm \,\sqrt{%
(E-\epsilon _n)^2-4}\right) 
\end{equation}
with $\lambda _n^{(1)}\lambda _n^{(2)}=1$. Therefore, for stronger disorder $%
W\gg 1\,$, one can compute the inverse localization length directly via the
largest value $\lambda _{+}$ of these two eigenvalues, by neglecting the
region $|E-\epsilon| < 2$, 
\begin{equation}
\label{l2} 
l_s^{-1}=\left\langle \ln \left| \lambda _{+}\right| \right\rangle =
\int \ln \left( \frac12 \left( x+ \sqrt{x^2-4}\right)\right) dx=
F(z_1)+F(z_2)
\end{equation}
Here, $x=|E-\epsilon |$ and $z_1=W/2+E, \, z_2=W/2-E$; the function $F$ is
defined by
\begin{equation}
\label{F}
F(z)= \frac 2W \left( z \ln \left( z+\sqrt{z^2-4} \right)-\sqrt{z^2-4}-z\ln 2
\right)
\end{equation}
Note that from this expression one can easily
obtain the known expression for the localization length in the limit of a
very strong disorder, 
\begin{equation}
\label{limit}l_s^{-1}=\ln \frac W2-1 
\end{equation}

\section{Concluding remarks}

We have shown in Section II how to derive exact expression for the
localization length in the weak disorder limit, using the properties of a
noisy circle map (\ref{smallmap}). In particular, at the center of the
energy band, a small correction to the localization length obtained by
standard perturbation theory is needed, which is due to the contribution of
the forth harmonics in the expression for the invariant measure $\rho
(\theta )$ . It is interesting that for other ``resonant'' values of the
energy, the (weak) modulation of $\rho (\theta )$ has no influence on the
localization length. However, for other quantities than the
localization length, these corrections may be important. In this sense, the
exact expression (\ref{stationary}) for the invariant measure 
$\rho (\theta)$ obtained
in this paper for a weak disorder, may find important applications.

We would like to point out that the calculation in Ref.~\cite{Der84} of the
localization length and of the density of states is performed after shifting
the energy $E\to E+x$ slightly away from the ``resonant'' values, the size
of the shift being proportional to the variance $x\sim \sigma ^2$ for all
energies except the band-edge. It is easy to see that the approach we have
used here, also allows us for the derivation of the localization length near
the center of the band. Moreover, we can also understand why the shift has
to be, as in Ref. \cite{Der84}, of the order of the variance of the disorder. In
fact, the recipe is that one should have $A_n\gg x$, because, if the shift
from a value of the energy corresponding to a rational value of $\alpha $ is
too large, the orbit becomes quasi-periodic. In this case the modulation of
the invariant measure which results in the non--trivial contribution to the
localization length,  is lost.

Finally, it is interesting to note that the 1D map (\ref{smallmap}) can be
compared with the Arnold map~\cite{Arn65} 
\begin{equation}
\theta _{n+1}=\theta _n-a+b\sin \,\theta
_n\,\,\,\,\,\,\,\,\,\,\,\,\,\,\,\,\,\,\,\,\,\,\mbox{mod}\,\,\,\,2\pi ~.
\end{equation}
If we approximate $A_n^2$ with its average, it is then tempting to associate
the parameter $a$ to our parameter $\mu $ , and the parameter $b$ to $%
\langle A_n^2\rangle $. Although the modulation of the circle map (\ref
{smallmap}) is a different function, and the noise is added through the term
containing $A_n$, our results show that the structure of the Arnold tongues
persists (Arnold tongues are regions of the parameter space \{$a,b$\} where
the dynamics is locked on a periodic orbit of period $q$, the tongues become
narrower and narrower as $b$ is reduced). Indeed, inside the tongue of the
Arnold map any orbit corresponds to a rational rotation number $p/q$;
outside the tongues the motion is quasi-periodic. Trajectories inside the
tongues of our model do display periodic motion with an additional
diffusion, the periodic motion being responsible for the modulation of the
invariant measure $\rho (\theta )$. Outside the tongue, the motion is
quasi-periodic also in our model and the invariant measure is flat.

\acknowledgements

We thank {\it Centro Internacional de Ciencias} in Cuernavaca, M\`exico,
where this work was finished for financial support and the Institute of
Scientific Interchange in Torino, where this work was started. F.M.I. thanks
the INFM-FORUM for funding his trips to Italy, as well as the support from
the INTAS grant No. 94-2058. This work is also part of the European contract
ERBCHRXCT940460 on ``Stability and universality in classical mechanics''.





\begin{thebibliography}{99}

\bibitem{Izr95a} F.M. Izrailev and S. Ruffo, unpublished.

\bibitem{Izr95b} F.M. Izrailev, T. Kottos, and G.P. Tsironis, \prb {\bf 52}, 
3274 (1995).

\bibitem{Izr97} T. Kottos, G.P. Tsironis and F.M. Izrailev, J. Phys.
Condens. Matter, {\bf 9}, 1777 (1997).

\bibitem{Kap81} M. Kappus and F. Wegner, Z. Phys. B {\bf 45}, 15 (1981).

\bibitem{Der84} B. Derrida and E. Gardner, J. Physique (Paris), {\bf 45},
1283 (1984).

\bibitem{Tho79} D.J. Thouless, in {\it Ill-condensed matter},
R. Balian, R. Maynard and G. Toulouse, eds., North Holland, Amsterdam
and NY (1979), p.1.

\bibitem{Czy81}
G. Czycholl, B. Kramer and A. MacKinnon, Z. Phys. B - Condensed Matter,
{\bf 43}, 5 (1981). 

\bibitem{Gardiner} C.W. Gardiner, {\it Handbook of Stochastic Methods},
Springer-Verlag, Berlin (1985).

\bibitem{Econo} E.N. Economou, {\it Green's Functions in Quantum Physics},
Springer-Verlag, Berlin (1984). 

\bibitem{Ben84} G. Benettin, Physica D {\bf 13}, 211 (1984).

\bibitem{Arn65} V. Arnold, Transl. Series 2, {\bf 46}, 213 (1965).

\end{thebibliography}
\end{document}